# Veryl: A New Hardware Description Language as an Alternative to SystemVerilog

Naoya Hatta, PEZY Computing, K.K., Tokyo, Japan (hatta@pezy.co.jp)

Taichi Ishitani, PEZY Computing, K.K., Tokyo, Japan (ishitani@pezy.co.jp)

Ryota Shioya, The University of Tokyo, Tokyo, Japan (shioya@ci.i.u-tokyo.ac.jp)

*Abstract*— **Veryl, a hardware description language based on SystemVerilog, offers optimized syntax tailored for logic design, ensuring synthesizability and simplifying common constructs. It prioritizes interoperability with SystemVerilog, allowing for smooth integration with existing projects while maintaining high readability. Additionally, Veryl includes a comprehensive set of development support tools, such as package managers and real-time checkers, to boost productivity and streamline the design process. These features empower designers to conduct high-quality hardware design efficiently.**

## I. Introduction

Hardware Description Languages (HDLs) are essential for designing digital circuits, with Verilog, VHDL, and SystemVerilog being the most widely used. Verilog and VHDL, originating in the 1980s, do not incorporate advanced language features that enhance abstraction and code reusability, unlike recent programming languages that adopt object-oriented and functional paradigms. SystemVerilog extends Verilog with advanced features; however, this complexity makes full support by electronic design automation (EDA) tools difficult and limits usability. Additionally, these languages blend constructs for logic design and simulation, necessitating careful selection by users. While modern programming languages provide productivity tools like auto-formatters and linters, traditional HDLs lack these supportive environments, making development challenging.

There is an approach to constructing HDLs as domain-specific languages (DSLs) within existing programming languages by extending their syntax and libraries. For example, Chisel [1], a DSL built on Scala, leverages the advanced abstraction features of Scala to create highly reusable HDLs, while MyHDL [2] extends Python. This approach benefits from the enhanced development tools and comprehensive libraries available for the base programming language.

However, the existing approaches have specific limitations. Primarily, constructing an HDL as a DSL within a programming language significantly constrains it by the syntax of the base language. This makes it difficult to provide specialized syntax for hardware description. For example, essential elements in HDLs, such as arbitrary bit-width representations, signal direction, and special signals like clock and reset, are not directly supported by the underlying programming language. Consequently, these elements must often be expressed redundantly through function calls or similar mechanisms.

Additionally, most existing approaches convert DSL code into Verilog code through compilers because many EDA tools support only traditional HDLs like Verilog and VHDL. However, due to the advanced abstraction capabilities of DSLs, this conversion often results in voluminous Verilog code that has significantly different semantics from the original DSL. This discrepancy makes the Verilog code difficult to read and debug. Moreover, ASIC design flows, such as timing closure and pre/post-mask engineering change order (ECO), frequently necessitate partial modifications to the Verilog code. Predicting the extent of necessary changes in the Verilog code from alterations in the DSL code is challenging, thereby complicating the application of these design flows.

To address the challenges, we are developing Veryl, a new hardware description language designed specifically for logic design. Veryl transpiles source code into highly readable SystemVerilog, ensuring robust interoperability with existing SystemVerilog codebases. It incorporates syntax elements from modern programming languages like Rust and Go, which enhance expressiveness and productivity. Features such as automatic formatting, linting, and real-time editor integration further boost its usability. In the subsequent sections, we will describe the features and the advantages of Veryl.

## II. Hardware Description Language Veryl

### A. Basic Syntax

The syntax of Veryl is based on SystemVerilog keywords and incorporates syntax improvements from modern programming languages. In Figure 1, the same module is described in both SystemVerilog and Veryl to highlight the basic syntactic differences. Marks such as '---①' are added for explanation and are not part of the source code.

```
// SystemVerilog code

// Counter
module Counter #(
    parameter WIDTH = 1
)(
    input  logic             i_clk  ,
    input  logic             i_rst_n,
    output logic [WIDTH-1:0] o_cnt
);
    logic [WIDTH-1:0] r_cnt;

    always_ff @ (posedge i_clk or negedge i_rst_n) begin
        if (!i_rst_n) begin
            r_cnt <= 0;
        end else begin
            r_cnt <= r_cnt + 1;
        end
    end

    always_comb begin
        o_cnt = r_cnt;
    end
endmodule
```

```
// Veryl code

/// Counter    -------------------------①
module Counter #(
    param WIDTH: u32 = 1,     -----------②
)(
    i_clk: input  clock       ,
    i_rst: input  reset       ,
    o_cnt: output logic<WIDTH>,    -----③
){
    var r_cnt: logic<WIDTH>;    --------④

    always_ff {    ---------------------⑤
        if_reset {
            r_cnt = 0;
        } else {
            r_cnt += 1;    -------------⑥
        }
    }

    always_comb {
        o_cnt = r_cnt;
    }
}
```

Figure 1. Basic code example.

- **Documentation Comment**: Comments that begin with '`///`' are considered documentation comments, as indicated at ①. These are syntactically distinct from regular comments, which start with '`//`', and are specifically used for generating documentation.
- **Trailing Comma**: Veryl supports a trailing comma after the last element in a comma-separated list, as shown at ②. This feature not only eliminates the need to adjust commas when adding or removing elements but also minimizes unnecessary differences in version control systems like Git.
- **Simple Array Syntax**: In SystemVerilog, the distinction between packed and unpacked arrays is made by the placement of '`[]`' either before or after the variable name. Veryl, however, uses '`<>`' for packed arrays and '`[]`' for unpacked arrays, allowing for direct width specification as illustrated at ③. This approach eliminates the redundant notation like '`[WIDTH-1:0]`' that is commonly found in SystemVerilog.
- **Type after Variable**: In Veryl, the type is specified after the variable name, as demonstrated at ④. This arrangement simplifies syntax parsing and supports abbreviated type notation by allowing omission of the details following the colon.

- **Abbreviated Clock and Reset**: In scenarios where there are only one clock and one reset within the scope, Veryl allows for the omission of clock and reset specifications in ‘`always_ff`’ declarations, as illustrated at ⑤. Detailed information about clock and reset handling will be covered in the next section.

- **Context-aware Assignment**: SystemVerilog differentiates between blocking and non-blocking assignments using ‘`=`’ and ‘`<=`’ operators, respectively. This distinction often leads to errors, such as using ‘`=`’ in ‘`always_ff`’ contexts causing unexpected synthesis result. Veryl simplifies this by using a single assignment operator ‘`=`’, where the nature of the assignment—blocking or non-blocking—is determined by the context, such as ‘`always_ff`’ or ‘`always_comb`’. This not only prevents common operator usage errors but also enables the use of compound assignment operators like ‘`+=`’ in non-blocking contexts, as shown at ⑥.

### *B. Clock and Reset*

Veryl introduces sophisticated syntax for clock and reset. Unlike SystemVerilog, which treats clock and reset as regular variables, Veryl classifies them as dedicated types, distinctly separating them from regular variables. As mentioned previously, this distinction enables the omission of clock and reset specifications in ‘`always_ff`’ declarations. Even in designs featuring multiple clocks and resets, modules typically operate with a single clock and reset, allowing these specifications to be omitted in such cases. This approach facilitates concise descriptions for most scenarios, while still providing the option to detail individual clocks and resets when necessary.

Furthermore, when implementing the same source code for both ASIC and FPGA targets, adjustments may be necessary for the polarity and synchronicity of the reset. Veryl facilitates this by allowing configuration of the polarity and synchronicity of clock and reset types during transpilation to SystemVerilog. In SystemVerilog, writing reset conditions might require expressions like ‘`if (!i_rst_n)`’ or ‘`if (i_rst)`’ depending on the reset polarity. In contrast, Veryl introduces a dedicated syntax for reset conditions, called ‘`if_reset`’, enabling the transpiler to automatically generate appropriate reset conditions. For scenarios where polarity and synchronicity are fixed, Veryl allows them to be explicitly defined using special types like ‘`reset_async_high`’, which denotes an asynchronous active-high reset.

Figure 2 demonstrates how different SystemVerilog code is generated from a single Veryl code by configuring the clock and reset settings.

```
// Veryl code
module ModuleA (
    i_clk_a: input `a clock           ,
    i_clk_b: input `b clock_negedge   ,
    i_rst_a: input `a reset           ,
    i_rst_b: input `b reset_async_high,
) {
    always_ff (i_clk_a, i_rst_a) {
        if_reset {
        }
    }
    always_ff (i_clk_b, i_rst_b) {
        if_reset {
        }
    }
}
```

```
// Generated SystemVerilog code with
//   clock_type = posedge
//   reset_type = async_low
always_ff @ (posedge i_clk_a or negedge i_rst_a) begin
    if (!i_rst_a) begin
    end
end
always_ff @ (negedge i_clk_b or posedge i_rst_b) begin
    if (i_rst_b) begin
    end
end

// Generated SystemVerilog code with
//   clock_type = negedge
//   reset_type = sync_high
always_ff @ (negedge i_clk_a) begin
    if (i_rst_a) begin
    end
end
always_ff @ (negedge i_clk_b or posedge i_rst_b) begin
    if (i_rst_b) begin
    end
end
```

Figure 2. Code example of clock and reset.

In the right upper code, the edge specification of 'i_clk_a' becomes 'posedge' because 'posedge' is specified as 'clock_type'. Similarly, the edge specification and reset condition of 'i_rst_a' are adjusted depending on 'async_low' which denotes asynchronous active-low reset. In the right lower code, 'i_rst_a' does not appear in the sensitivity list because the specified reset type is 'sync_high', indicating synchronous active-high reset. The generated code of 'i_clk_b' and 'i_rst_b' is not affected by clock and reset configuration because the type of them has the specified polarity and synchronicity.

Additionally, clock domain annotation enables the identification of the clock domain to which each signal belongs. The annotation is represented by identifier with a single quotation mark, such as '`a'. In the code shown on the left of Figure 2, 'i_clk_a' and 'i_rst_a' belong to the same clock domain, '`a', while 'i_clk_b' and 'i_rst_b' belong to the '`b' domain. This annotation allows the Veryl compiler to detect unexpected clock domain crossings. The expected clock domain crossing points are specified by the 'unsafe (cdc)' block, enabling reviewers to focus their checks on these specific points.

### C. *Generics*

SystemVerilog allows module and interface customization through parameter overrides during instantiation. However, this customization is limited to elements that can be specified as parameters, such as values, but not to elements like the names of modules to be instantiated. To address this limitation, Veryl introduces generics, enabling more versatile descriptions of modules, interfaces, packages, functions, and structures.

For example, consider the following code, which demonstrates a queue utilizing SRAM for storage. SRAM module naming conventions vary across different intellectual property (IP) vendors or technology processes, requiring SystemVerilog to duplicate code for each SRAM name or replace names through text macros. In contrast, Veryl allows the insertion of the module name from the outside, decoupling the logic of the queue from the SRAM module name. This approach significantly enhances code reusability.

```
module SramQueue::<T> {
    inst u_sram: T;

    // queue logic
}

module Test {
    // Instantiate a SramQueue by SramVendorA
    inst u0_queue: SramQueue::<SramVendorA>();

    // Instantiate a SramQueue by SramVendorB
    inst u1_queue: SramQueue::<SramVendorB>();
}
```

Figure 3. Code example of generics.

## III. VERYL COMPILER AND DEVELOPMENT ENVIRONMENT

Veryl is designed not only to advance as a hardware description language, as previously mentioned, but also to boost productivity throughout the compiler and the entire development environment. In this section, we will explore several key features offered by the Veryl compiler.

### A. *Semantic Check and Real-time Diagnostics*

The Veryl compiler not only generates SystemVerilog code as a transpiler but also performs a range of semantic checks to ensure code quality. Here are some examples of the checks it provides:

- Detects duplication and undefined identifiers.
- Checks for multiple assignments, uninitialized variables, and unused variables.
- Prevents the unintended generation of latches.
- Ensures consistency between signal direction and assignment/reference.
- Ensures that module ports and function arguments are consistent with their invocation.
- Detects bit-width overflow in numerical literals.

Additionally, the Veryl compiler includes a language server that integrates with various text editors like Visual Studio Code [3] and Vim [4]. Operating under the Language Server Protocol [5], established by Microsoft, this server communicates with the editor to provide real-time feedback during the editing process as illustrated in Figure 4. This feedback includes the results of semantic checks, which can correct errors typically identified by costly linting tools or during the logic synthesis stage in traditional SystemVerilog workflows.

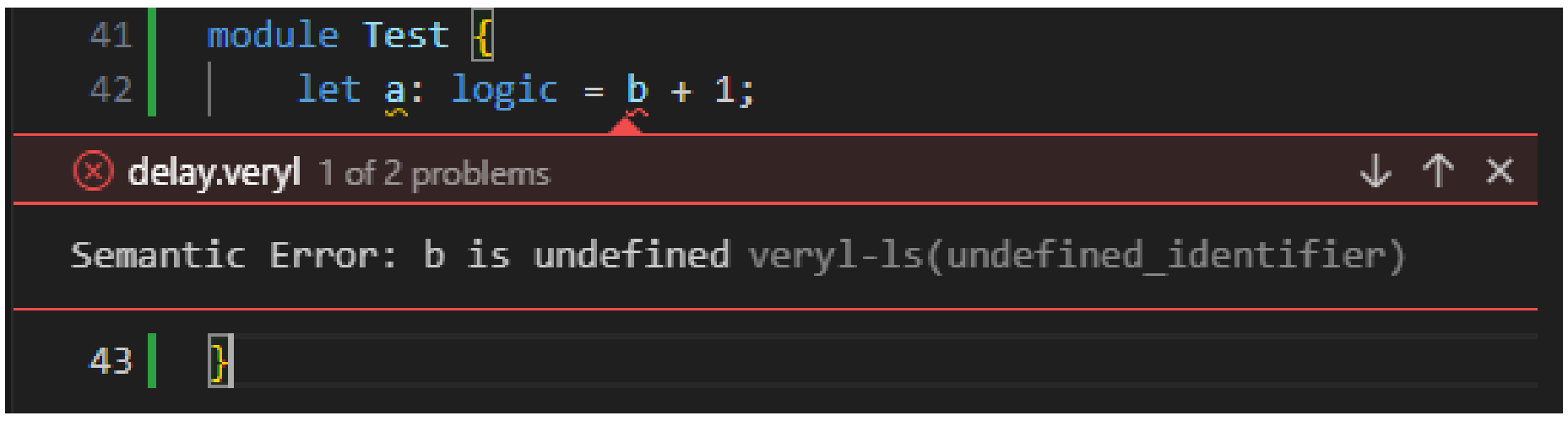

Figure 4. Real-time diagnostics.

*B. Interoperability with SystemVerilog*

User-defined types in Veryl, such as modules, interfaces, packages, and structures, are fully equivalent to those in SystemVerilog, enabling mutual referencing. This compatibility facilitates the gradual integration of Veryl into existing projects by either combining it with existing SystemVerilog code or by rewriting parts of those projects in Veryl.

Furthermore, with debuggers and waveform viewers that support SystemVerilog, developers can directly interact with Veryl-defined types, which enhances debugging efficiency.

Additionally, the correspondence of variables and '`always_ff`' declaration between Veryl code and the generated SystemVerilog code ensures seamless integration into ASIC workflows, including timing closure and ECO.

*C. Library Support*

In recent years, the practice of utilizing a vast array of open-source libraries to build large-scale software has become prevalent among programming languages. Similarly, Veryl incorporates features that facilitate the efficient use of existing libraries, enabling developers to seamlessly integrate diverse resources into their projects.

For example, developers can seamlessly integrate external libraries into their Veryl projects by simply adding the specified entry to the project definition file, as demonstrated in Figure 5. This entry imports the designated Git repository, enabling the utilization of modules, interfaces, and packages defined within that library.

```
[dependencies]
"https://github.com/veryl-lang/sample" = "0.1.0"
```

Figure 5. Configuration example to use library.

```
/// This is a sample module.
///
/// ```wavedrom
/// {signal: [
///   {name: 'i_clk', wave: 'p......'},
///   {name: 'i_dat', wave: 'x.=x...', data: ['data']},
///   {name: 'o_dat', wave: 'x...=x.', data: ['data']},
/// ]}
/// ```
pub module Sample #(
    /// Data Width
    param WIDTH: u32 = 1,
) (
    i_clk: input  clock       , /// Clock
    i_dat: input  logic<WIDTH>, /// Input Data
    o_dat: output logic<WIDTH>, /// Output Data
) {}
```

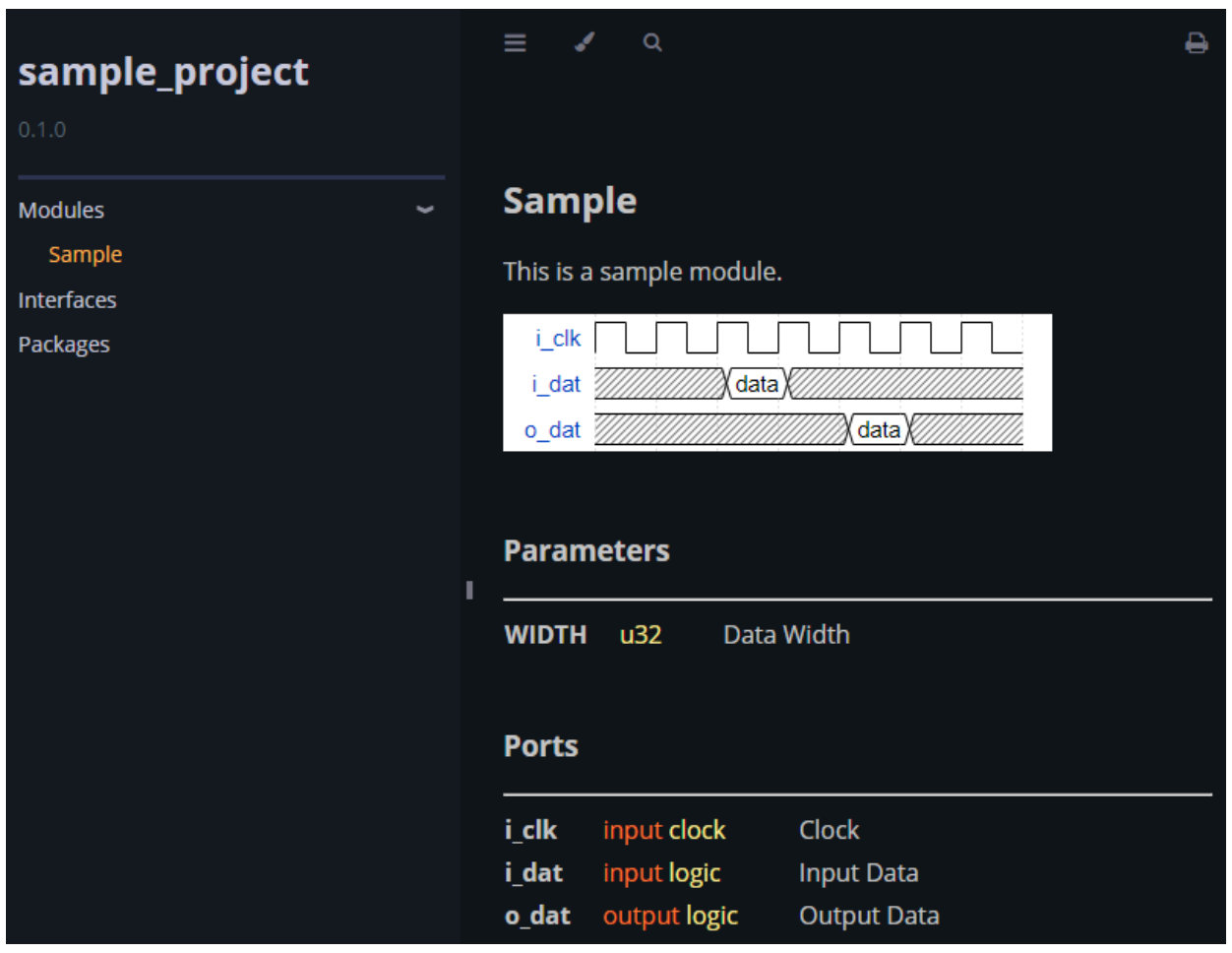

Figure 6. Documentation generation.

Furthermore, documentation is a crucial element when using libraries. Typically, when each library author chooses their preferred location and format for documentation, users must individually search for and extract necessary information from each library, which can be cumbersome. In Veryl, however, the compiler supports automatic documentation generation. The documentation comments within the source code are compatible with CommonMark [6] syntax and can include waveforms in WaveDrom [7] format. The Veryl compiler interprets these comments, extracts information about modules, interfaces, and packages, and automatically generates consistently formatted documentation, as illustrated in Figure 6 from the provided source code.

## IV. Conclusion

Veryl is a new hardware description language tailored for logic design, featuring specialized syntax and high interoperability with SystemVerilog. It also offers comprehensive compiler and development environment support to enhance productivity. Developed as open-source software, Veryl is publicly available at https://github.com/veryl-lang/veryl, with the aim of achieving widespread adoption.

## Acknowledgment

Thank you to everyone who participated in discussions about the language and tool design at https://github.com/veryl-lang/veryl, and to those who contributed to the source code.